\documentclass[reprint,superscriptaddress,amsmath,amssymb,aps,floatfix]{revtex4-1}
\usepackage{graphicx}
\usepackage{dcolumn}
\usepackage{upgreek}
\usepackage{gensymb}
\usepackage{float}
\usepackage{bm}
\usepackage{color}

\begin{document}


\title{Highly efficient spin orbit torque in Pt/Co/Ir multilayers with antiferromagnetic interlayer exchange coupling}


\author{Yuto Ishikuro}
\affiliation{Department of Physics, The University of Tokyo, Tokyo 113-0033, Japan}

\author{Masashi Kawaguchi}
\affiliation{Department of Physics, The University of Tokyo, Tokyo 113-0033, Japan}

\author{Tomohiro Taniguchi}
\affiliation{National Institute of Advanced Industrial Science and Technology (AIST), Spintronics Research Center, Tsukuba 305-8568, Japan}

\author{Masamitsu Hayashi}
\affiliation{Department of Physics, The University of Tokyo, Tokyo 113-0033, Japan}
\affiliation{National Institute for Materials Science, Tsukuba 305-0047, Japan}


\newif\iffigure
\figurefalse
\figuretrue

\date{\today}

\begin{abstract}
We have studied the spin orbit torque (SOT) in Pt/Co/Ir multilayers with 3 repeats of the unit structure. As the system exhibits oscillatory interlayer exchange coupling (IEC) with varying Ir layer thickness, we compare the SOT of films when the Co layers are coupled ferromagnetically and antiferromagnetically.  
SOT is evaluated using current induced shift of the anomalous Hall resistance hysteresis loops. A relatively thick Pt layer, serving as a seed layer to the multilayer, is used to generate spin current via the spin Hall effect. In the absence of antiferromagnetic coupling, the SOT is constant against the applied current density and the corresponding spin torque efficiency (i.e. the effective spin Hall angle) is $\sim$0.09, in agreement with previous reports. In contrast, for films with antiferromagnetic coupling, the SOT increases with the applied current density and eventually saturates. The SOT at saturation is a factor of $\sim$15 larger than that without the antiferromagnetic coupling. The spin torque efficiency is $\sim$5 times larger if we assume the net total magnetization is reduced by a factor of 3 due to the antiferromagnetic coupling. Model calculations based on the Landau Lifshitz Gilbert equation show that the presence of antiferromagnetic coupling can increase the SOT but the degree of enhancement is limited, in this case, to a factor of 1.2-1.4.
We thus consider there are other sources of SOT, possibly at the interfaces, which may account for the highly efficient SOT in the uncompensated synthetic anti-ferromagnet (SAF) multilayers. 
\end{abstract}


\maketitle
\section{Introduction}
Spin orbit torque (SOT)\cite{manchon2009prb} is considered as a viable means to manipulate magnetization of thin magnetic layers for next generation magnetic random access memories (MRAM)\cite{garello2018vlsi}. Bilayers consisting of a non-magnetic metal (NM) and a ferromagnetic metal (FM) are widely used as a prototype\cite{miron2011nature,liu2012science} to demonstrate the feasibility of SOT technologies. The spin torque efficiency is often defined as a parameter that characterizes both the degree of spin current generated from the NM layer and the effectiveness of the spin current to exert spin torque on the magnetic moments of the FM layer. For a spin transparent NM/FM interface, the spin torque efficiency is equivalent to the spin Hall angle of the NM layer. 

To improve the spin torque efficiency, significant effort has been put forward to explore materials with large spin Hall angle.  Beyond the 5\textit{d} transition metals, recent studies have shown that topological insulators\cite{mellnik2014nature,fan2014nmat}, van der Waals materials\cite{macneill2017nphys} and antiferromagnets exhibit large spin torque efficiency.  
In particular, antiferromagnetic materials are attracting interest as an efficient spin current source which are readily accessible\cite{zhang2014prl,zhang2016sciadv}. 
Recent experiments have demonstrated current controlled magnetization switching of ferromagnetic layer using antiferromagnetic thin films as the spin current source\cite{fukami2016nmat}. In addition to the conventional intrinsic and extrinsic spin Hall effects, antiferromagnetic materials may have additional means to generate spin current due to their unique magnetic structure\cite{chen2014prl,zhang2017prb}. The large anomalous Hall effect\cite{nakatsuji2015nature,nayak2016sciadv} and the spin Hall effect\cite{zhang2016sciadv} in chiral antiferromagnets are known as a consequence of electrons acquiring Berry's phase as they travel through the magnetic texture. 

Collinear antiferromagnets\cite{duine2018nphys} can be designed by means of interlayer exchange coupling (IEC) of thin ferromagnetic layers\cite{parkin1990prl}. As the net magnetic moment can be reduced to near zero, such synthetic anti-ferromagnets (SAF) play an essential role in modern MRAM technologies: they are typically used as the magnetic reference layer owing to their negligible stray field\cite{parkin1999jap}. The small net magnetization is also attractive with regard to their use as the information recording layer (i.e. free layer). As the current needed to control the magnetization direction of the free layer scales with its saturation magnetization $M_\mathrm{S}$, smaller $M_\mathrm{S}$ is desirable for low power operation provided that one can keep the thermal stability factor sufficiently high. Recent studies have shown that the efficiency to manipulate the magnetization direction of ferrimagnets or synthetic antiferromagnets using spin orbit torques can be significantly increased when the net magnetization, or the net angular momentum, of such magnets is reduced to near zero\cite{roschewsky2016apl,finley2016prap,mishra2017prl,ueda2017prb,zhang2018prb,krishnia2019jmmm}. Similarly, the velocity of magnetic domain walls driven by field\cite{kim2017nmat} or current\cite{yang2015nnano,caretta2018nnano} can be enhanced when the net angular momentum or magnetization is minimized.

Here we compare spin orbit torque switching of Pt/Co/Ir multilayers with ferromagnetic and antiferromagnetic interlayer exchange coupling. We use multilayers consisting of three repeats of the unit structure Pt/Co/Ir: the film is an uncompensated SAF with a non-zero net magnetization if the three Co layers are coupled antiferromagnetically. A relatively thick Pt layer, serving as a seed layer to the multilayer, is used to generate spin current via the spin Hall effect. The SOT of the uncompensated SAF is nearly 15 times larger than that when the Co layers are coupled ferromagnetically. The spin torque efficiency is $\sim$5 times larger for the former if we assume the net total magnetization is 3 times smaller for the antiferromagnetically coupled state, although the dominant SOT may be exerted at the bottom Co layer in contact with the Pt seed layer. We model the system to study possible mechanisms that can cause such large enhancement of SOT due to antiferromagnetic IEC. 


\section{Experimental setup}
Multilayers composed of Sub./3 Ta/2 Pt/[0.6 Pt/0.9 Co/$d_\mathrm{Ir}$ Ir]$_{3\times}$/2 MgO/1 Ta (units in nanometer) were grown on Si substrates with SiO$_2$ coating (thickness: 100 nm) using radio frequency magnetron sputtering. The thickness of the Ir layer ($\sim$0.1 nm$\leq d_\mathrm{Ir} \leq\sim$1.1 nm) was varied across the substrate using a moving shutter during the deposition process. 
Optical lithography and Ar ion milling were used to form a Hall-bar. Schematic of the experimental setup and definition of the coordinate axis are shown in Fig.~\ref{fig:Rxy}(b). The width of the wire and the distance between the longitudinal voltage probes are $\sim$10 $\mu$m and $\sim$25 $\mu$m, respectively. DC current ($I$) was passed along the $x$ axis. Positive current is defined as current flow to $+x$. The Hall resistance ($R_{xy}$) is obtained by dividing the measured Hall voltage ($V_{xy}$) with the current supplied, i.e. $R_{xy}=V_{xy}/I$. 

\section{Results and discussion}
\subsection{Film characteristics}
\begin{figure*}[t]
	\includegraphics[scale=0.6]{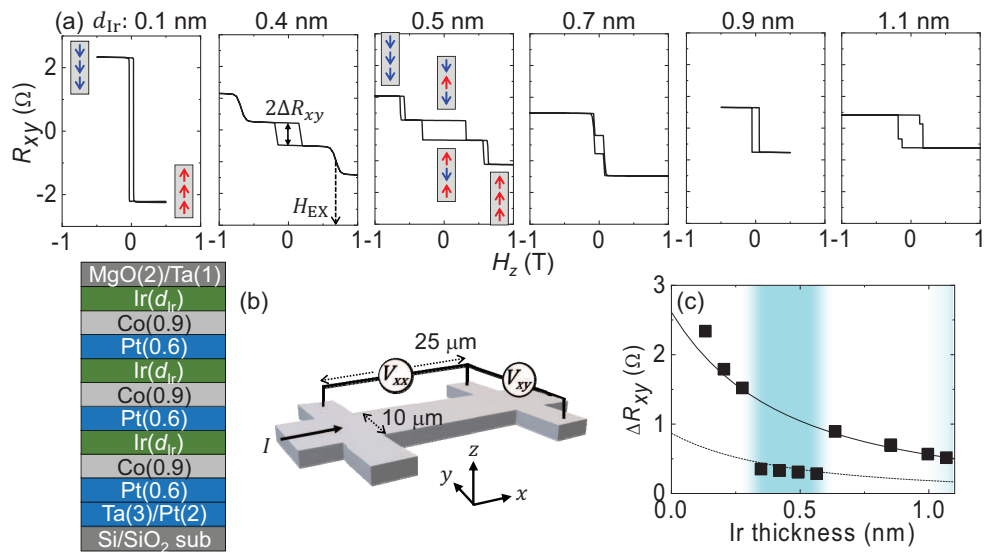}
	\caption{(a) Hall resistance ($R_{xy}$) plotted as a function of out of plane field $H_z$. From left to right: the thickness of the Ir layer ($d_\mathrm{Ir}$) is varied as indicated above each panel. The arrows in the gray box indicate the corresponding magnetization state of $R_{xy}$: the top, middle and bottom arrows represent the magnetization direction of the top, middle and bottom Co layers in the multilayer. Definitions of $\Delta R_{xy}$ and $H_\mathrm{EX}$ are sketched. (b) Schematic illustration of the film stacking, Hall bar and the measurement setup. The coordinate axis employed is sketched. (c) The anomalous Hall resistance ($\Delta R_{xy}$) at zero field plotted as a function of Ir layer thickness. The solid and dotted lines show the calculated $\Delta R_{xy}$ using Eq.~(\ref{eq:Rxy}). The parameters used are: $\rho_\mathrm{Co}=35$ $\mu \Omega$cm, $\rho_\mathrm{Pt}=40$ $\mu \Omega$cm, $\rho_\mathrm{Ir}=15$ $\mu \Omega$cm\cite{kawaguchi2018apl,ishikuro2019prb}. $\theta_\mathrm{AH}\sim0.1$ for the solid line and $\theta_\mathrm{AH}\sim0.033$ for the dotted line. The blue shaded regions represent states with antiferromagnetic coupling.
	\label{fig:Rxy}}
\end{figure*}

Figure~\ref{fig:Rxy}(a) shows the out-of-plane field ($H_z$) dependence of the Hall resistance ($R_{xy}$) for films with various Ir layer thicknesses.  In this thickness range (0.1 nm $\leq d_\mathrm{Ir} \leq$ 1.1 nm), oscillatory interlayer exchange coupling is observed: the coupling is either ferromagnetic (F) or antiferromagnetic (AF). For the films with F coupling, the hysteresis loops show two stable states corresponding to the three Co layers' magnetization all pointing along $+z$ (from top to bottom, the Co layers' magnetization are $\uparrow$$\uparrow$$\uparrow$) and -z ($\downarrow$$\downarrow$$\downarrow$). For such coupling, the three Co layers switch together at the same field. In contrast, we typically find four stable states for the films with AF coupling: 
the four states correspond to two saturated states ($\uparrow$$\uparrow$$\uparrow$ and $\downarrow$$\downarrow$$\downarrow$) and two intermediate states with the middle Co layer magnetization pointing against that of the two neighboring layers ($\uparrow$$\downarrow$$\uparrow$ and $\downarrow$$\uparrow$$\downarrow$). See the arrows displayed in Fig.~\ref{fig:Rxy}(a) for the corresponding magnetization configuration of the four $R_{xy}$ states. The field at which switching between the parallel to antiparallel states occurs is defined as $H_\mathrm{EX}$, as schematically defined in Fig.~\ref{fig:Rxy}(a). Note that the hysteresis loop of the film with $d_\mathrm{Ir}$$\sim$1.1 nm indicates that the AF coupling is in place but weak such that the field range which the antiparallel states appear is small. 

Figure 1(c) shows the Ir layer thickness ($d_\mathrm{Ir}$) dependence of the anomalous Hall resistance $\Delta R_{xy}$. $
\Delta R_{xy}$ is defined as the difference of $R_{xy}$ for the two metastable states at zero-field, i.e. at remanence. $\Delta R_{xy}$ decreases with increasing $d_\mathrm{Ir}$ due to current shunting into the Ir layer. The blue shaded regions in Fig.~\ref{fig:Rxy}(c) display the Ir thickness range in which the AF coupled state is stable. To study the magnetic configuration of the films at remanence via the anomalous Hall resistance, we model the transport properties of the heterostructure assuming current flow within the highly conducting Pt, Co and Ir layers. Current flow into the Ta seed layer, which has a significantly larger resistivity than the conducting layers, is neglected. Note that the MgO/Ta capping layer does not conduct current (the top Ta layer is oxidized and forms an insulator). We define the resistivities (thickness) of the Pt, Co and Ir layers as, $\rho_\mathrm{Pt} (d_\mathrm{Pt}), \rho_\mathrm{Co} (t_\mathrm{Co}), \rho_\mathrm{Ir} (d_\mathrm{Ir})$, respectively. Assuming a parallel circuit model, $\Delta R_{xy}$ reads 
\begin{equation}
\label{eq:Rxy}
\Delta R_{xy}=\tan\theta_\mathrm{AH} \frac{\rho_\mathrm{Co}}{t_\mathrm{Co}} \big( 1 + \frac{\rho_\mathrm{Co} d_\mathrm{N}}{\rho_\mathrm{N} t_\mathrm{Co}} \big)^{-2}
\end{equation}
where $\frac{d_\mathrm{N}}{\rho_\mathrm{N}} \equiv \frac{d_\mathrm{Pt}}{\rho_\mathrm{Pt}} + \frac{d_\mathrm{Ir}}{\rho_\mathrm{Ir}}$ and $\theta_\mathrm{AH}$ is the effective anomalous Hall angle. We use Eq.~(\ref{eq:Rxy}) to characterize the results of the films with F- and AF-coupling.
Assuming that the resistivity of the conducting layers is the same for the films with F- and AF-coupling, the difference in $\Delta R_{xy}$ between the two can be attributed to the net magnetization along the $z$ axis, which is implicitly included in $\theta_\mathrm{AH}$. The solid and dotted lines in Fig.~\ref{fig:Rxy}(c) show the calculated $\Delta R_{xy}$, with $\theta_\mathrm{AH}$ of the dotted line being 1/3 of that of the solid line.
These results are consistent with the picture that the remanent state of the films with AF coupling have net magnetization that is one third of that of the F coupling films.

\subsection{Current induced torque}
\begin{figure}[t]
	\includegraphics[scale=0.5]{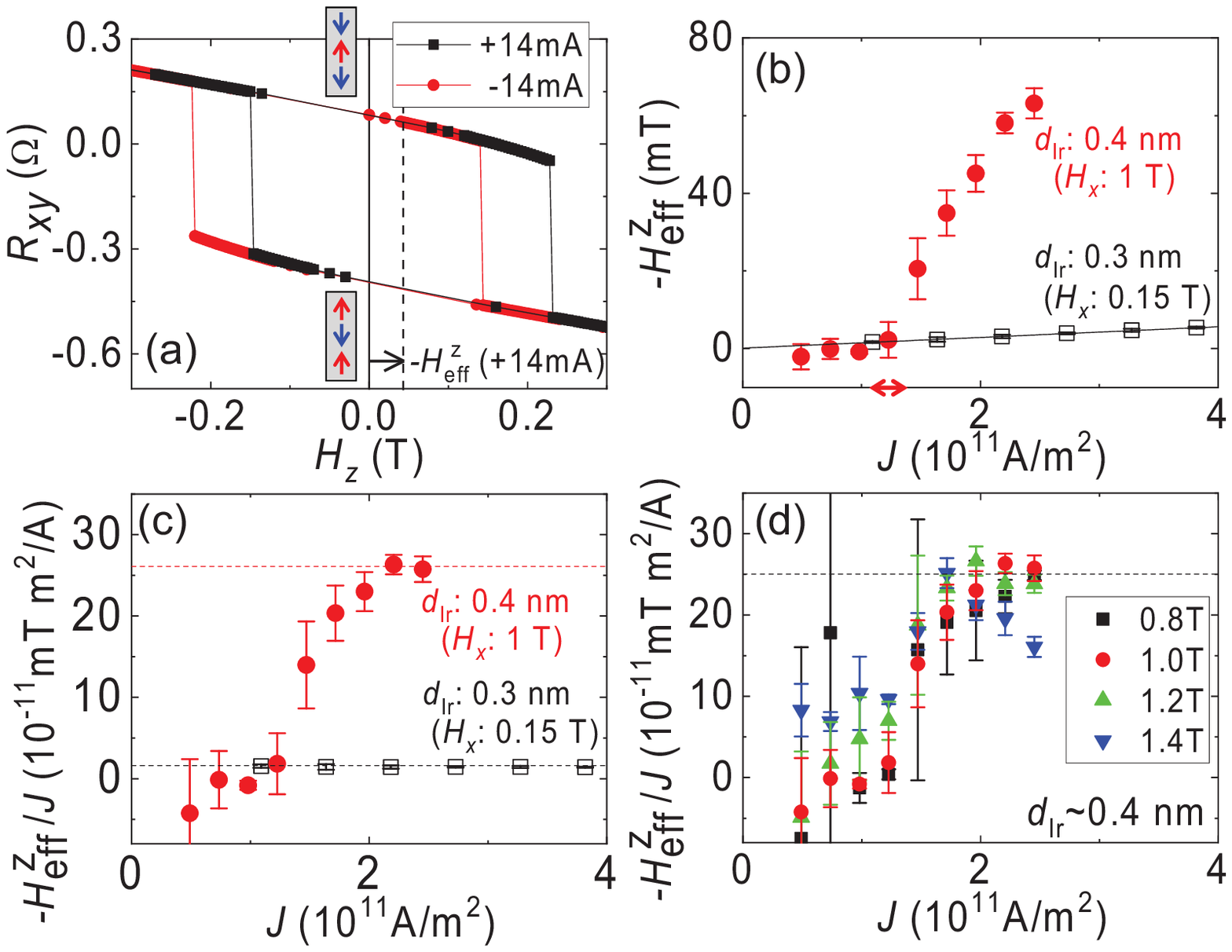}
	\caption{(a) The Hall resistance ($R_{xy}$) vs. $H_z$ when DC current of 14 mA (black squares) or -14 mA (red circles) is applied to the Hall bar made from a film with $d_\mathrm{Ir} \sim 0.4$ nm. A constant in-plane field ($H_x \sim 1$ T) along the $x$-axis is applied during the measurements. Definition of $H_\mathrm{eff}^z$ and the magnetic configuration of the multilayer near zero field are schematically illustrated. (b) The current density ($J$) dependence of $H_\mathrm{eff}^z$ for $d_\mathrm{Ir} \sim 0.3$ nm (ferromagnetic coupling, open black squares) and $d_\mathrm{Ir} \sim 0.4$ nm (antiferromagnetic coupling, solid red circles). A linear fit to the data with $d_\mathrm{Ir} \sim 0.3$ nm is shown by the black solid line. The threshold current density for the antiferromagnetically coupled state is schematically sketched using the red arrow. The applied $H_x$ is noted in the legend. (c) $H_\mathrm{eff}^z/J$ vs $J$ obtained from (b). (d) $H_\mathrm{eff}^z/J$ vs $J$ for the film with $d_\mathrm{Ir} \sim 0.4$ nm. $H_x$ is varied as denoted. The error bars in (b) represent standard deviation of repeated measurements of $H_\mathrm{eff}^z$ and those in (c) are the errors carried over from (b). Similar definition also applies to the error bars in (d).
	\label{fig:switching}}
\end{figure}

The current-induced shift of the hysteresis loops were used to estimate the spin torque efficiency of the multilayers\cite{pai2016prb}. A constant bias field directed along the current flow ($H_x$) was applied while the Hall resistance was measured as a function of $H_z$. Figure 2(a) shows exemplary $R_{xy}$-$H_z$ loops for a film with AF coupling ($d_\mathrm{Ir} \sim$0.4 nm) when positive and negative currents were applied under a bias field $H_x$$\sim$1 T. The two metastable states at zero field represent the antiferromagnetically coupled states $\uparrow$$\downarrow$$\uparrow$ and $\downarrow$$\uparrow$$\downarrow$. When positive (negative) current is applied, the center of the hysteresis loop shifts to positive (negative) $H_z$. The shift of the loop center with respect to $H_z$=0 is defined as $-H_\mathrm{eff}^z$\cite{pai2016prb}. In the AF coupled films, we note that the current-induced shift of the hysteresis loops is nearly zero for the switching between the saturated states and the antiparallel states, i.e. transitions from $\uparrow$$\uparrow$$\uparrow$ to $\uparrow$$\downarrow$$\uparrow$ states, $\downarrow$$\downarrow$$\downarrow$ to $\downarrow$$\uparrow$$\downarrow$ states, and vice versa. Note that these switching takes place at larger $|H_z|$ compared to that between the antiparallel states ($\uparrow$$\downarrow$$\uparrow$ to $\downarrow$$\uparrow$$\downarrow$ and vice versa). We therefore focus on the switching between the $\uparrow$$\downarrow$$\uparrow$ and the $\downarrow$$\uparrow$$\downarrow$ states.

Figure 2(b) shows the current density ($J$) dependence of $H_\mathrm{eff}^z$ for films with AF coupling ($d_\mathrm{Ir} \sim$0.4 nm) and F coupling ($d_\mathrm{Ir} \sim$0.3 nm). The in-plane bias field was set to $H_x$$\sim$1 T for the former and $H_x$$\sim$0.15 T for the latter. $J$ is obtained by dividing the current ($I$) with the width of the wire and the total thickness of the conducting Pt, Co and Ir layers: although the resistivities of the Pt, Co and Ir layers are different, we assume a uniform current flow within these layers for simplicity. As evident in Fig.~\ref{fig:switching}(b), $H_\mathrm{eff}^z$ increases linearly with $J$ for the film with F coupling (open squares). In contrast, for the film with AF coupling (solid circles), $H_\mathrm{eff}^z$ increases abruptly above a threshold current density of $J \sim1.3\times10^{11}$ A/m$^2$. To obtain the spin torque efficiency, it is customary to divide $H_\mathrm{eff}^z$ with $J$\cite{pai2016prb}. Figure 2(c) displays $H_\mathrm{eff}^z/J$ as a function of $J$. $H_\mathrm{eff}^z/J$ is constant for all $J$ for the film with F coupling (open squares) whereas it saturates at $J\sim2.1\times10^{11}$ A/m$^2$ for the film with AF coupling (solid circles). Interestingly, the saturated value of $H_\mathrm{eff}^z/J$ for the latter (film with AF coupling) is significantly larger than the constant $H_\mathrm{eff}^z/J$ of the former (F coupling).

As the in-plane bias field ($H_x$) applied during the hysteresis loop measurements is different for the films with F and AF couplings, we study the $H_x$ dependence of $H_\mathrm{eff}^z/J$. For single magnetic layer films (e.g. NM/FM bilayers), it is known that $H_\mathrm{eff}^z/J$ takes a constant value when the magnitude of $H_x$ is larger than that of the Dzyaloshinskii-Moriya (DM) exchange field $H_{\textrm{DM}}$: the constant $H_\mathrm{eff}^z/J$ for $|H_x| > | H_{\textrm{DM}} |$ is proportional to the spin torque efficiency $\xi_{\textrm{DL}}$\cite{pai2016prb}. This is also the case for films with multiple magnetic layers coupled ferromagnetically\cite{ishikuro2019prb}. For multilayer films with AF coupling, here we show in Fig.~\ref{fig:switching}(d) $H_\mathrm{eff}^z/J$ vs $J$ obtained using different $H_x$ for the film with $d_\mathrm{Ir} \sim0.4$ nm. Although values of $H_\mathrm{eff}^z/J$ varies with $H_x$ when $J$ is smaller than the threshold current density, the saturated value of $H_\mathrm{eff}^z/J$ (above $J\sim2 \times 10^{11}$ A/m$^2$) is almost the same within the applied field ($H_x$) range. We thus take the $H_\mathrm{eff}^z/J$ upon saturation, defined as $h_\mathrm{eff}^z$ hereafter, as a measure of the spin torque efficiency for the films with AF coupling. For the films with F coupling, we assign $h_\mathrm{eff}^z$ as the constant $H_\mathrm{eff}^z/J$ when $|H_x| > | H_{\textrm{DM}} |$.

Note that the current-induced shift of the hysteresis loops for the films with AF coupling becomes near zero when $H_x$$\sim$0.4 T, suggesting that $H_{\textrm{DM}}$ lies between 0.4 T and 0.8 T ($d_\mathrm{Ir} \sim0.4$ nm). This is significantly larger than the $H_{\textrm{DM}}$ of the multilayer films with F coupling reported previously ($H_{\textrm{DM}} \sim 0.08$ T)\cite{ishikuro2019prb}.
For the films with AF coupling, we consider the field $H_x$ required to cause saturation of $H_\mathrm{eff}^z/J$, which has been assigned as $H_{\textrm{DM}}$ previously, is related to the emergence of IEC.
Experimentally, such saturation field correspond to $H_x$ needed to align the magnetization direction of all domain walls. 
If IEC is present in the system, the exchange coupling field $H_\mathrm{EX}$ acts on the domain walls, and thus it will take extra field to align their magnetization direction along $H_x$. As $H_\mathrm{EX}$ is $\sim0.6$ T to $\sim0.7$ T, it is likely that the saturation field is dominated by $H_\mathrm{EX}$.

\begin{figure}[t]
	\includegraphics[scale=0.5]{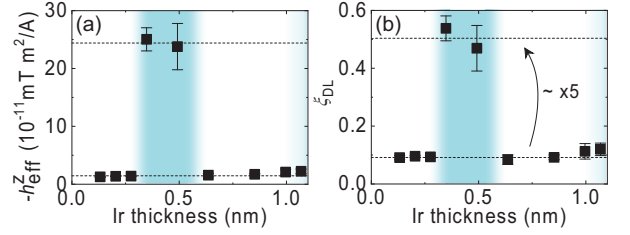}
	\caption{(a) The saturation value of $H_\mathrm{eff}^z/J$ for films with AF coupling, and the constant value of $H_\mathrm{eff}^z/J$ for the films with F coupling, both denoted as $h_\mathrm{eff}^z$, plotted as a function of the Ir layer thickness $d_\mathrm{Ir}$. (b) $d_\mathrm{Ir}$ dependence of the spin torque efficiency $\xi_{\textrm{DL}}$. The blue shaded regions in (a,b) represent states with antiferromagnetic coupling. The average values of $h_\mathrm{eff}^z$ and $\xi_{\textrm{DL}}$ for the ferromagnetic and antiferromagnetic coupled states are shown by the horizontal broken lines. The error bars are associated with the standard deviation of repeated measurements of $H_\mathrm{eff}^z$. For the films with AF coupling, the error bars are obtained from $H_\mathrm{eff}^z/J$ upon saturation.
	\label{fig:SOT}}
\end{figure}

The Ir layer thickness dependence of $h_\mathrm{eff}^z$ is plotted in Fig.~\ref{fig:SOT}(a). The average $h_\mathrm{eff}^z$ for films with F (AF) coupling is $\sim$1.5 mT ($\sim$25 mT) per current density of $10^{11}$ A/m$^2$. To convert $h_\mathrm{eff}^z$ to the spin torque efficiency ($\xi_{\textrm{DL}}$), we use the relation (for $H_x>0$) 
\begin{equation}
\label{eq:xiDL}
\xi_{\textrm{DL}} = - \frac{4 e M_\mathrm{eff} t_\mathrm{F}}{\pi \hbar} h_\mathrm{eff}^z,
\end{equation}
where $e (>0)$ is the electric charge, $\hbar$ is the reduced Planck constant, and $t_\mathrm{F}$ is the total thickness of the magnetic layers.
$M_\mathrm{eff}$ is the effective saturation magnetization of the system. 
Since the net magnetic moment of the films with AF coupling is one third of that of the films with F coupling, we substitute $M_\mathrm{eff}=M_\mathrm{S} / 3$ for the former and $M_\mathrm{eff}=M_\mathrm{S}$ for the latter in Eq.~(\ref{eq:xiDL}). $M_\mathrm{S}$ is the saturation magnetization of the Co layer which is estimated using magnetometry measurements. 
As $M_\mathrm{S}$ slightly varies with $d_\mathrm{Ir}$\cite{ishikuro2019prb}, we interpolate the data to obtain the $M_\mathrm{S}$ for the Hall bars studied here.
Note that the dominant SOT likely takes place at the bottom Co layer which is in contact with the relatively thick Pt seed layer. It is not obvious if the spin current impinging on the bottom Co layer will see a three times reduced magnetization when the three Co layers are coupled antiferromagnetically. The use of $M_\mathrm{eff}=M_\mathrm{S} / 3$ for the films with AF coupling thus remains as an issue that need to be addressed.
$\xi_{\textrm{DL}}$ is plotted as a function of $d_\mathrm{Ir}$ in Fig.~\ref{fig:SOT}(b). 
As evident, $\xi_{\textrm{DL}}$ for the films with AF coupling is nearly five times larger than that of the films with F coupling, reaching a value of $\sim$0.5.

\subsection{Model calculations}
Previously it has been reported that the spin torque efficiency can be enhanced by using antiferromagnetically coupled magnetic layers\cite{mishra2017prl,zhang2018prb}. In particular, similar value of $H_\mathrm{eff}^z/J$ was reported in a completely compensated synthetic antiferromagnet, which was associated with the nearly zero net magnetic moment\cite{zhang2018prb}. Here we find that the spin torque efficiency ($\xi_{\textrm{DL}}$) is large even though the net magnetic moment is not zero. 
To account for these results, we study the effect of the so-called exchange coupling torque\cite{yang2015nnano,mishra2017prl} on $\xi_{\textrm{DL}}$.

We first consider a NM/FM bilayer to which both the SOT and the external magnetic field are applied. Dynamics of the magnetization can be described using the Landau-Lifshitz-Gilbert equation:
\begin{equation}
\label{eq:llg}
\frac{d \bm{m}}{dt} = -\gamma \bm{m} \times \bm{H}_{\textrm{EFF}} + \alpha \bm{m} \times \frac{d \bm{m}}{dt} - \gamma H_\mathrm{DL} \bm{m} \times (\bm{m} \times \bm{p}),
\end{equation}
where $\bm{m}$ is the magnetization unit vector and $\alpha$ is the Gilbert damping constant of the magnetic layer. 
$\gamma$ is the gyromagnetic ratio, $\bm{H}_{\textrm{EFF}}$ is the effective magnetic field that acts on $\bm{m}$. The effective field can be expressed as $\bm{H}_{\textrm{EFF}} = -\frac{1}{M_\mathrm{S}} \frac{\delta E}{\delta \bm{m}}$, where $M_\mathrm{S}$ and $E$ are the saturation magnetization and the total energy of the system. $\frac{\delta f}{\delta \bm{m}}$ is meant to take the functional derivative of $f$ with $\bm{m}$. The energy density takes the form 
\begin{equation}
\label{eq:energy}
\frac{E}{V}= - M_\mathrm{S} \bm{m} \cdot \bm{H}_{\textrm{ext}} - K_{\textrm{eff}} m_z^2,
\end{equation}
where $V$ is the volume of the magnetic layer, $K_{\textrm{eff}} = \frac{1}{2} M_\mathrm{S} H_{\mathrm{K}0}$ is the uniaxial perpendicular magnetic anisotropy energy density. $H_{\mathrm{K}0}$ and $\bm{H}_{\textrm{ext}}$ are the anisotropy field and the external field, respectively. 
The third term on the right hand side of Eq.~(\ref{eq:llg}) represents the spin orbit torque on $\bm{m}$. $\bm{p}$ is a unit vector that represents the polarization of the spin current that diffuses into the magnetic layer and $H_\mathrm{DL}$ is the damping-like spin orbit effective field.
We assume the current flows in the NM layer along the $x$ axis and generates a spin current with $\bm{p} = (0, -1, 0)$ that diffuses into the magnetic layer (here we set $\bm{p}$ such that it agrees with the spin Hall effect of Pt and the stacking order, i.e. FM layer deposited on NM (Pt) layer).
 
We look for a solution at equilibrium when the current is small. At equilibrium, we substitute $\frac{d \bm{m}}{dt} = 0$ in Eq.~(\ref{eq:llg}) to obtain
\begin{equation}
\label{eq:equilibrium}
  \bm{m}
  \times
  \left(
    \bm{H}_\mathrm{EFF}
    +
    H_\mathrm{DL} 
    \bm{m}
    \times
    \bm{p}
  \right)
  =
  \bm{0}.
\end{equation}
In accordance with experiments, we apply an in-plane field along the current: $\bm{H}_{\textrm{ext}} =  (H_x, 0, 0)$. For simplicity, we neglect the $z$ component of the field. Substituting these parameters into Eq.~(\ref{eq:equilibrium}), we obtain
\begin{equation}
\label{eq:equilibrium2}
  \begin{pmatrix}
    (H_{\mathrm{K}0} m_{z}-H_\mathrm{DL}m_{x})m_{y} \\
    (H_x+H_\mathrm{DL}m_{z})m_{z}-(H_{\mathrm{K}0}m_{z}-H_\mathrm{DL}m_{x})m_{x} \\
    -(H_x+H_\mathrm{DL}m_{z})m_{y}
  \end{pmatrix}
  =
  \bm{0}.
\end{equation}
Under application of small current, the magnetization direction is set by the anisotropy and external fields. $\bm{m}$ thus lies in the $zx$ plane, i.e. $m_y \approx 0$. We therefore substitute $m_y=0$ into Eq.~(\ref{eq:equilibrium2}) to obtain
\begin{equation}
\label{eq:equilibrium3}
  \left(
    H_x
    +
    H_\mathrm{DL}
    m_{z}
  \right)
  m_{z}
  -
  \left(
    H_{\mathrm{K}0}
    m_{z}
    -
    H_\mathrm{DL}
    m_{x}
  \right)
  m_{x}
  =
  0.
\end{equation}
From hereon, we express $\bm{m}$ using the spherical coordinates, i.e. $\bm{m}=(\sin\theta\cos\varphi,\sin\theta\sin\varphi,\cos\theta)$. ($\varphi =0$ for $m_y=0$.)
Without current ($H_\mathrm{DL} = 0$), we obtain from Eq.~(\ref{eq:equilibrium3}), $m_x = H_x/H_{\mathrm{K}0}$. Assuming $H_x \ll H_{\mathrm{K}0}$, we obtain $m_x \sim \theta$ and find
\begin{equation}
\label{eq:equilibrium4_noI}
  \theta (I = 0)
  \simeq
  \frac{H_x}{H_{\mathrm{K}0}}.
\end{equation}
Under the application of current, we use linear approximation and drop higher order terms of $m_x$ to obtain
\begin{equation}
\label{eq:equilibrium4_I}
  \theta (I \neq 0)
  \simeq
  \frac{H_x + H_\mathrm{DL}}{H_{\mathrm{K}0}}.
\end{equation}
The difference in $\theta$ with and without current reads
\begin{equation}
\label{eq:dtheta}
  \Delta \theta
  \equiv
  \theta (I \neq 0)
  -
  \theta (I = 0)
  =
  \frac{H_\mathrm{DL}}{H_{\mathrm{K}0}}.
\end{equation}

Experimentally, the spin orbit effective field is evaluated using the following formula.
\begin{equation}
\label{eq:HDLexp}
H_\mathrm{DL}^\mathrm{exp} = \frac{\Delta \theta}{\frac{\partial \theta}{\partial H_x}}.
\end{equation}
Combining Eq.~(\ref{eq:equilibrium4_noI}), from which we find $\frac{\partial \theta}{\partial H_x} = \frac{1}{H_{\mathrm{K}0}}$, and Eqs.~(\ref{eq:dtheta}) and (\ref{eq:HDLexp}), we obtain 
\begin{equation}
\label{eq:HDLexp_single}
H_\mathrm{DL}^\mathrm{exp} = H_\mathrm{DL},
\end{equation}
which is what we expect for the single layer system.

Next we consider two magnetic layers A and B coupled antiferromagnetically. We assume the spin orbit torque only acts on layer A. The LLG equations of unit magnetization vector $\bm{m}_{A(B)}$ of layer A (B) are
\begin{equation}
\begin{aligned}
\label{eq:llg_AB}
\frac{d \bm{m}_A}{dt} =& -\gamma \bm{m}_A \times \bm{H}_{\textrm{EFF,}A} + \alpha_A \bm{m}_A \times \frac{d \bm{m}_A}{dt}\\
 &\ \ \ \ \ \ - \gamma H_\mathrm{DL} \bm{m}_A \times (\bm{m}_A \times \bm{p}),\\
\frac{d \bm{m}_B}{dt} =& -\gamma \bm{m}_B \times \bm{H}_{\textrm{EFF,}B} + \alpha_B \bm{m}_B \times \frac{d \bm{m}_B}{dt},
\end{aligned}
\end{equation}
where $\bm{H}_{\textrm{EFF,}l} = -\frac{1}{M_l} \frac{\delta E}{\delta \bm{m}_l}$ ($l=A,B$).
The total areal energy density of the system reads 
\begin{equation}
\begin{aligned}  
\label{eq:energy_AB}
  \frac{E}{S}
  &=
  -M_{A} \bm{H}_{\textrm{ext}} \cdot \bm{m}_{A} t_{A}
  -
  M_{B} \bm{H}_{\textrm{ext}} \cdot \bm{m}_{B} t_{B}\\
  &-
  K_{\textrm{eff},A} m_{Az}^{2} t_{A}
  -
  K_{\textrm{eff},B} m_{Bz}^{2} t_{B}
  -
  J_\mathrm{EX}
  \bm{m}_{A}
  \cdot
  \bm{m}_{B},
  \end{aligned}
\end{equation}
where $M_{A(B)}$, $H_{\textrm{K},A(B)}$, $\alpha_{A(B)}$ and $t_{A(B)}$ represent the saturation magnetization, the anisotropy field, the Gilbert damping constant and the thickness of layer A (B), respectively. The uniaxial perpendicular magnetic anisotropy energy density of layer $A(B)$ is defined as $K_{\textrm{eff},A(B)} = \frac{1}{2} M_{A(B)} H_{\textrm{K}0,A(B)}$. $J_\mathrm{EX}$ is the interlayer exchange coupling constant and $S$ is the area of interface between layers A and B. Negative $J_\mathrm{EX}$ stabilizes antiferromagnetic IEC.
Similar to the experimental setup, we assume the two layers A and B are composed of the same material with the same thickness. We therefore set $M_A = M_B = M_\mathrm{S}$, $H_{\textrm{K}0,A} = H_{\textrm{K}0,B} = H_{\mathrm{K}0}$, $\alpha_A = \alpha_B = \alpha$ and $t_A = t_B = t$. We define the exchange coupling field 
\begin{equation}
\label{eq:H_EX}
H_\mathrm{J} \equiv - \frac{J_\mathrm{EX}}{M_\mathrm{S} t}.
\end{equation}
Again, we look for the equilibrium state under small current. Substituting $\frac{d \bm{m}_l}{dt} = 0$ and $\bm{H}_{\textrm{ext}} =  (H_x, 0, 0)$ into Eq.~(\ref{eq:llg_AB}) and using $m_{Ay}=m_{By}=0$, we obtain
\begin{equation}
\begin{aligned}
\label{eq:torque_A}
  &\left(
    H_x
    -
    H_\mathrm{J}
    m_{Bx}
    +
    H_\mathrm{DL}
    m_{Az}
  \right)
  m_{Az} \\
  &= 
  \left(
    H_{\mathrm{K}0}
    m_{Az}
    -
    H_\mathrm{J}
    m_{Bz}
    -
    H_\mathrm{DL}
    m_{Ax}
  \right)
  m_{Ax},
\end{aligned}
\end{equation}
from the first equation of Eq.~(\ref{eq:llg_AB}) and 
\begin{equation}
\begin{aligned}
\label{eq:torque_B}
  \left(
    H_x
    -
    H_\mathrm{J}
    m_{Ax}
  \right)
  m_{Bz}
  =
  \left(
    H_{\mathrm{K}0}
    m_{Bz}
    -
    H_\mathrm{J}
    m_{Az}
  \right)
  m_{Bx},
\end{aligned}
\end{equation}
from the second equation. We express the magnetization vectors using spherical coordinates: $\bm{m}_A = (\sin\theta_A\cos\varphi_A,\sin\theta_A\sin\varphi_A,\cos\theta_A)$ and $\bm{m}_B = (\sin\theta_B\cos\varphi_B,\sin\theta_B\sin\varphi_B,\cos\theta_B)$. With $H_x \ll H_{\mathrm{K}0}$, we drop higher order terms of $m_{Ax}$ and $m_{Bx}$. We assume $\theta_A \ll 1$ and, due to the antiferromagnetic exchange coupling, $\theta_B = \pi - \theta_B^{\prime}$ with $\theta_B^{\prime} \ll 1$. 
Substituting these relations into Eqs.~(\ref{eq:torque_A}) and (\ref{eq:torque_B}), we obtain the form of polar angle of layers A and B when the current is turned off ($H_\mathrm{DL} = 0$) as
\begin{equation}
\begin{aligned}
\label{eq:theta_AB_noI}
  \theta_{A} (I = 0)
  &\simeq
  \frac{H_x}{H_{\mathrm{K}0}+2H_\mathrm{J}},
\\
  \theta_{B}^{\prime} (I = 0)
  &\simeq
  \frac{H_x}{H_{\mathrm{K}0}+2H_\mathrm{J}}.
\end{aligned}
\end{equation}
Turning on the current, we find
\begin{equation}
\begin{aligned}
\label{eq:theta_AB_I}
  \theta_{A} (I \neq 0)
  &\simeq
  \frac{H_x H_{\mathrm{K}0}+H_\mathrm{DL}(H_{\mathrm{K}0}+H_\mathrm{J})}{H_{\mathrm{K}0}(H_{\mathrm{K}0}+2H_\mathrm{J})},
\\
  \theta_{B}^{\prime} (I \neq 0)
  &\simeq
  \frac{H_xH_{\mathrm{K}0}-H_\mathrm{DL}H_\mathrm{J}}{H_{\mathrm{K}0}(H_{\mathrm{K}0}+2H_\mathrm{J})}.
\end{aligned}
\end{equation}
The difference in the polar angle with and without current therefore reads
\begin{equation}
\begin{aligned}
\label{eq:dtheta_AB}
  \Delta \theta_A \equiv \theta_A (I \neq 0) - \theta_A (I = 0)
  &=
  \frac{\frac{H_{\mathrm{K}0}+H_\mathrm{J}}{H_{\mathrm{K}0}} H_\mathrm{DL}}{H_{\mathrm{K}}},
\\
 \Delta \theta_B^{\prime} \equiv \theta_B^{\prime} (I \neq 0) - \theta_B^{\prime} (I = 0)
  &=
   -\frac{\frac{H_\mathrm{J}}{H_{\mathrm{K}0}}H_\mathrm{DL}}{H_{\mathrm{K}}},
\end{aligned}
\end{equation}
where we have defined the effective anisotropy field when the exchange coupling field is non-zero:
\begin{equation}
\begin{aligned}
\label{eq:HKeff}
H_\mathrm{K} \equiv H_{\mathrm{K}0}+2H_\mathrm{J}.
\end{aligned}
\end{equation}
This definition follows from Eqs.~(\ref{eq:equilibrium4_noI}) and (\ref{eq:theta_AB_noI}).
Note that $H_\mathrm{K}$ corresponds to the experimentally measured anisotropy field under the influence of antiferromagnetic interlayer exchange coupling\cite{knepper2005prb,lau2019prm}.
(For the films with F coupling, $H_\mathrm{J}=0$ and $H_\mathrm{K} = H_{\mathrm{K}0}$.)

Again, we use Eq.~(\ref{eq:HDLexp}) to obtain the spin orbit effective field:
\begin{equation}
\label{eq:HDLexpAB}
H_{\mathrm{DL},l}^\mathrm{exp} = \frac{\Delta \theta_l}{\frac{\partial \theta_l}{\partial H_x}},
\end{equation}
where $H_{\mathrm{DL},l}^\mathrm{exp}$ is the spin orbit effective field that acts on layer $l=A,B$.
From Eq.~(\ref{eq:theta_AB_noI}), $\frac{\partial \theta}{\partial H_x}$ can be calculated. 
Substituting the results and Eq.~(\ref{eq:dtheta_AB}) into Eq.~(\ref{eq:HDLexpAB}), we obtain
\begin{equation}
\begin{aligned}
\label{eq:HDL_AB}
H_{\textrm{DL,}A}^\mathrm{exp} &= \frac{H_{\mathrm{K}0}+H_\mathrm{J}}{H_{\mathrm{K}0}} H_\mathrm{DL}, \\
H_{\textrm{DL,}B}^\mathrm{exp} &= -\frac{H_\mathrm{J}}{H_{\mathrm{K}0}} H_\mathrm{DL}.
\end{aligned}
\end{equation}
For both layers A and b, $H_\mathrm{J}$ increases the spin orbit effective field\cite{mishra2017prl}. 
In the limit of $H_\mathrm{J} \rightarrow \infty$, where the two layers act as a single FM layer, $H_{\textrm{DL,}A}^\mathrm{exp}$ and $H_{\textrm{DL,}B}^\mathrm{exp}$ diverge. 
Experimentally, however, what is being probed is $\Delta \theta_l$ ($l=A, B$). In the single layer limit (i.e., NM/FM bilayer), $\Delta \theta = \frac{H_\mathrm{DL}}{H_{\mathrm{K}0}}$ (see Eq.~(\ref{eq:dtheta})).
For two layers with antiferromagnetic coupling, according to Eq.~(\ref{eq:dtheta_AB}), $\Delta \theta_A \rightarrow \frac{H_\mathrm{DL}}{2 H_{\mathrm{K}0}}$ and $\Delta \theta_B \rightarrow - \frac{H_\mathrm{DL}}{2 H_{\mathrm{K}0}}$ as $H_\mathrm{J} \rightarrow \infty$. Aside from the factor of 2 in the denominator, which is caused by the assumption that spin current only acts on layer A, the two systems return the same results in the limit of $H_\mathrm{J} \rightarrow \infty$.


Although $H_{\textrm{DL,}l}^\mathrm{exp}$ ($l=A,B$) increases with increasing $H_\mathrm{J}$, it should be noted that the effective magnetic anisotropy field $H_\mathrm{K}$ also increases with $H_\mathrm{J}$. 
Thus the efficiency of the spin orbit effective field, characterized by $\Delta \theta_l$ ($l=A,B)$, does not necessarily increase with the strength of IEC. 
As we discuss in the next section, the spin orbit effective field found in the multilayers with AF coupling is significantly larger than what we expect from Eq.~(\ref{eq:HDL_AB}).
Under such circumstance, the efficiency $\Delta \theta_l$ ($l=A,B)$ can be significantly larger than the case without the AF coupling. 

\subsection{Evaluation of the exchange coupling torque}

These results show that the spin orbit effective field that acts on the magnetization of each layer increases with increasing strength of IEC. As the spin current diffuses into layer A in this model, the effective field is always larger for layer A. 
We therefore consider $H_{\mathrm{DL},A}^\mathrm{expt}$ provides an upper limit of the spin orbit effective field for the multilayer system. To compare experimental results with the model calculations, we focus on the relative size of the spin-orbit effective field ($h_\mathrm{eff}^z$ in the experiments and $H_{\mathrm{DL},A}^\mathrm{expt}$ in the model) with and without the antiferromagnetic coupling.
To estimate the degree of enhancement of the SOT due to the IEC, we first estimate $H_{\mathrm{K}0}$ and $H_J$. In the Appendix, Fig.~\ref{fig:HK}(b), we show the $d_\mathrm{Ir}$ dependence of $H_\mathrm{K}$ measured using transport measurements. For the films with F coupling, $H_\mathrm{K} = H_{\mathrm{K}0} \sim 1.5$ T when $d_\mathrm{Ir} \sim 0.6-0.9$ nm. 
The reduction of $H_\mathrm{K}$ for the thinner Ir films ($d_\mathrm{Ir} \sim 0.1-0.2$ nm) may be caused by non-uniform thickness of the Ir layer.
The measured $H_\mathrm{K}$ for the films with AF coupling is $\sim$2.6 T ($d_\mathrm{Ir} \sim 0.4$ nm) and $\sim$2.1 T ($d_\mathrm{Ir} \sim 0.5$ nm). Assuming the films with AF coupling have the same $H_{\mathrm{K}0}$ with that of the F coupling films ($H_{\mathrm{K}0} \sim 1.5$ T), we estimate, using Eq.~(\ref{eq:HKeff}), $H_\mathrm{J} \sim 0.55$ T ($d_\mathrm{Ir} \sim 0.4$ nm) and $H_\mathrm{J}\sim$0.3 T ($d_\mathrm{Ir} \sim 0.5$ nm).
We may compare these values to what we obtain from the switching field $H_\mathrm{EX}$ between the parallel and antiparallel magnetization states. $H_\mathrm{EX}$ estimated from the results shown in Fig.~\ref{fig:Rxy} give $H_\mathrm{EX}\sim0.7$ T ($d_\mathrm{Ir} \sim 0.4$ nm) and $H_\mathrm{EX}\sim0.6$ T ($d_\mathrm{Ir} \sim 0.5$ nm). For an antiferromagnetically coupled two FM layer system, $H_\mathrm{EX}$ can be obtained analytically\cite{lau2019prm}:
\begin{equation}
\begin{aligned}
H_{\textrm{EX}} \approx \frac{1}{2} \big(\sqrt{H_{\textrm{K}0}^2 + 2 H_{\textrm{K}0} H_{\textrm{J}}} - (H_{\textrm{K}0} - 2 H_{\textrm{J}}) \big).
\label{eq:Hex}
\end{aligned}
\end{equation}
Although the samples evaluated here consist of three FM layers coupled antiferromagnetically, we may use Eq.~(\ref{eq:Hex}) as a first order approximation to characterize the experimentally obtained $H_\mathrm{EX}$.
Substituting $H_{\textrm{K}0}$ and $H_\mathrm{EX}$ into Eq.~(\ref{eq:Hex}), we obtain $H_\mathrm{J} \sim 0.5$ T ($d_\mathrm{Ir} \sim 0.4$ nm) and $\sim$0.4 T ($d_\mathrm{Ir} \sim 0.5$ nm), which are in good agreement with those estimated from $H_\mathrm{K}$. 

Substituting these values ($H_\mathrm{J} \sim 0.55$ T, $H_{\mathrm{K}0} \sim 1.5$ T for the film with $d_\mathrm{Ir} \sim 0.4$ nm, $H_\mathrm{J} \sim 0.4$ T, $H_{\mathrm{K}0} \sim 1.5$ T for the film with $d_\mathrm{Ir} \sim 0.5$ nm) into Eq.~(\ref{eq:HDL_AB}), we find $H_{\textrm{DL,}A}^\mathrm{exp} \sim 1.4 H_\mathrm{DL}$ for $d_\mathrm{Ir}\sim0.4$ nm and $H_{\textrm{DL,}A}^\mathrm{exp} \sim 1.2 H_\mathrm{DL}$ for $d_\mathrm{Ir}\sim0.5$ nm. 
Thus this model itself cannot account for the factor of 15 increase of $H_\mathrm{eff}^z$ when the Co layers are coupled antiferromagnetically. 
Note that the effective anisotropy field (Eq.~(\ref{eq:HKeff})) increases by a factor of $\sim$1.5 to $\sim$1.7 when the antiferromagnetic coupling is in place compared to that without it.
Thus the difference in the experimentally obtained $h_\mathrm{eff}^z$ for films with AF and F couplings (i.e. a factor of 15) is significantly larger for what the model predicts (Eq.~(\ref{eq:dtheta_AB}).

We therefore infer that there are other sources of SOT that may account for the highly efficient SOT acting on synthetic antiferromagnetic layers. 
Recent studies have revealed that SOT may originate from interface states\cite{amin2018prl,baek2018nmat} and spin currents from the ferromagnetic layer\cite{taniguchi2015prap,iihama2018nelec,amin2019prb,wang2019nnano}.  
In multilayer systems\cite{jamali2013prl}, it has been reported that the SOT increases with the number of repeats of the unit structure\cite{huang2015apl,jinnai2017apl}. 
We infer that the antiferromagnetically coupled magnetic states can create spin dependent electron potential well\cite{yuasa2002science} within the multilayers that influences spin transport and consequently the SOT\cite{stiles2002prb}. 
Further investigation, including spin transport modeling, is required to clarify the origin of the SOT in multilayers with antiferromagnetically coupled magnetic layers.


\section{Conclusion}

In conclusion, we have studied spin orbit torque switching of antiferromagnetically coupled Pt/Co/Ir multilayers. We use multilayers with three repeats of the unit structure. The interlayer exchange coupling varies with the Ir layer thickness.
When the Co layers are coupled antiferromagnetically, the system is an uncompensated synthetic antiferromagnet (SAF) with the net total magnetization three times smaller than that of the multilayer with ferromagnetic coupling. 
A relatively thick Pt seed layer is used as a source of spin current via the spin Hall effect of Pt. The spin orbit effective field is studied using current induced shift of the easy axis magnetic hysteresis loop obtained from the anomalous Hall resistance measurements.

We find the damping-like effective field of the uncompensated SAF is nearly 15 times larger than that of the multilayers with ferromagnetic coupling. The spin torque efficiency, which depends on the saturation magnetization of the ferromagnetic layer, is $\sim$5 times larger for the uncompensated SAF if we consider the net total magnetization, which is 3 times smaller for the SAF, is responsible for the SOT. 
Model calculations show that the antiferromagnetic interlayer exchange coupling can enhance the SOT.
The enhancement is the strongest for the Co layer that is in contact with the Pt seed layer. However the enhancement factor is limited to $\sim$1.2-1.4, which is considerably smaller than the factor of 15 we find experimentally. 
We thus infer that there are other effects that cause the highly efficient SOT for the synthetic antiferromagnetic multilayers:  for example, the spin dependent electron potential well that develops for antiferromagnetically coupled magnetic state can influence spin transport and may generate interface SOT that enhances the overall torque.

\begin{acknowledgments}
Acknowledgments: This work was partly supported by JSPS Grant-in-Aid for Specially Promoted Research (15H05702) and the Center of Spintronics Research Network of Japan.
\end{acknowledgments}

\appendix
\section{Magnetic properties of the multilayers}
\begin{figure}[h]
	\includegraphics[scale=0.5]{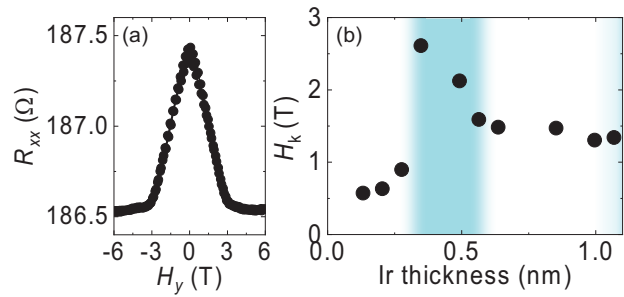}
	\caption{(a) Longitudinal resistance $R_{xx}$ plotted as a function of $H_y$ for a Hall bar made from a film with $d_\mathrm{Ir} \sim 0.4$ nm. The field at which $R_{xx}$ saturates is defined as $H_\mathrm{K}$. (b) The Ir layer thickness $d_\mathrm{Ir}$ dependence of the anisotropy field $H_\mathrm{K}$. The blue shaded regions in (b) represent states with antiferromagnetic coupling. 
	\label{fig:HK}}
\end{figure}

We use transport measurements to evaluate the anisotropy field of the multilayers. The longitudinal resistance $R_{xx}$ of the Hall bar is measured as a function of in-plane magnetic field $H_y$ orthogonal to the current flow (along the $y$ axis). Typical plot of $R_{xx}$ vs. $H_y$ from a multilayer with antiferromagnetic coupling is shown in Fig.~\ref{fig:HK}(a). $R_{xx}$ drops as $H_y$ is increased from zero due to the spin Hall magnetoresistance (SMR)\cite{nakayama2013prl,chen2013prb,kim2016prl}. The field at which $R_{xx}$ saturates correspond to the the anisotropy field $H_\mathrm{K}$. The Ir layer thickness dependence of $H_\mathrm{K}$ is plotted in Fig.~\ref{fig:HK}(b). The results are similar to those reported in Ref.~\cite{lau2019prm}. Note that the seed layer in this study is 3 Ta/2 Pt whereas it was 1 Ta/3 Ru in Ref.~\cite{lau2019prm}.

\newpage
\bibliography{reference_112419}

\end{document}